\documentclass[prl,twocolumn,superscriptaddress,amsmath,amssymb]{revtex4}

\usepackage[T1]{fontenc}
\usepackage{dcolumn}
\usepackage{bm}
\usepackage{epsf,color,graphicx}
\DeclareGraphicsExtensions{.jpg, .png , .pdf, .ps}

\begin{document}

\title{$\Pi$-stacking functionalization through micelles swelling:  Application to the synthesis of single wall carbon nanotube/porphyrin complexes for energy transfer}
\author{Cyrielle Roquelet}
\affiliation{Laboratoire de Photonique Quantique et
Mol\'eculaire, Institut d'Alembert, CNRS, ENS Cachan, 61 avenue
du pr\'esident Wilson, 94235 Cachan Cedex, France}
\author{Jean-Sébastien Lauret}
\email{lauret@lpqm.ens-cachan.fr} \affiliation{Laboratoire de
Photonique Quantique et Mol\'eculaire, Institut d'Alembert, CNRS,
ENS Cachan, 61 avenue du pr\'esident Wilson, 94235 Cachan Cedex,
France}
\author{Valérie Alain-Rizzo}
\affiliation{Laboratoire de Photophysique et Photochimie
Supramol\'eculaires et Macromol\'eculaires, Institut d'Alembert,
CNRS, ENS Cachan, 61 avenue du pr\'esident Wilson, 94235 Cachan
Cedex, France}
\author{Christophe Voisin}
\affiliation{Laboratoire Pierre Aigrain, Ecole Normale
Sup\'erieure, UPMC, Universit\'e Denis Diderot, CNRS, 24 rue
Lhomond, 75005 Paris, France}
\author{Romain Fleurier}
\affiliation{Laboratoire d'Etude des Microstructures, ONERA,
CNRS, 29 avenue de la division Leclerc, 92322 Chatillon, France}
\author{Morgan Delarue}
\affiliation{Laboratoire de Photonique Quantique et
Mol\'eculaire, Institut d'Alembert, CNRS, ENS Cachan, 61 avenue
du pr\'esident Wilson, 94235 Cachan Cedex, France}
\author{Damien Garrot}
\affiliation{Laboratoire de Photonique Quantique et
Mol\'eculaire, Institut d'Alembert, CNRS, ENS Cachan, 61 avenue
du pr\'esident Wilson, 94235 Cachan Cedex, France}
\author{Annick Loiseau}
\affiliation{Laboratoire d'Etude des Microstructures, ONERA,
CNRS, 29 avenue de la division Leclerc, 92322 Chatillon, France}
\author{Philippe Roussignol}
\affiliation{Laboratoire Pierre Aigrain, Ecole Normale
Sup\'erieure, UPMC, Universit\'e Denis Diderot, CNRS, 24 rue
Lhomond, 75005 Paris, France}
\author{Jacques A. Delaire}
\affiliation{Laboratoire de Photophysique et Photochimie
Supramol\'eculaires et Macromol\'eculaires, Institut d'Alembert,
CNRS, ENS Cachan, 61 avenue du pr\'esident Wilson, 94235 Cachan
Cedex, France}
\author{Emmanuelle Deleporte}
\affiliation{Laboratoire de Photonique Quantique et
Mol\'eculaire, Institut d'Alembert, CNRS, ENS Cachan, 61 avenue
du pr\'esident Wilson, 94235 Cachan Cedex, France}
\date{\today}

\begin{abstract}
We report on a new, orginal and efficient method for
"$\pi$-stacking" functionalization of single wall carbon
nanotubes. This method is applied to the synthesis of a
high-yield light-harvesting system combining single wall carbon
nanotubes and porphyrin molecules. We developed a micelle
swelling technique that leads to controlled and stable complexes
presenting an efficient energy transfer. We demonstrate the key
role of the organic solvent in the functionalization mechanism.
By swelling the micelles, the solvent helps the non water soluble
porphyrins to reach the micelle core and allows a strong
enhancement of the interaction between porphyrins and nanotubes.
This technique opens new avenues for the functionalization of
carbon nanostructures.
\end{abstract}

\maketitle

Molecular engineering of functionalized nano-materials is the
object of intensive research in connection to a wide range of
applications \cite{Nel2009}. Such complexes aim at combining the
unique properties of nano-objects and the wide tunability of
organic molecules properties. Especially, complexes presenting
charge or energy transfer are intensively investigated for biology
\cite{Taraska2009,Nel2009,Sacca2009} or optoelectronics
\cite{Hardin2009,Ehli2009,Park2009,Ross2009} applications. For
instance, fluorescence energy transfer (FRET) has been used to
monitor structural rearrangements in proteins \cite{Taraska2009}
or to enhance the excitation transfer in dye-sensitized solar
cells \cite{Hardin2009}. In this context, due to their unique
transport properties and especially their large carrier mobility,
single-wall carbon nanotubes (SWNTs) are of particular interest
for use in organic photovolta\"{\i}c devices
\cite{Campidelli2008,Ehli2009,ehli06,alvaro06,hasobe06,Guldi2008,Prato2009}.

Excitation transfer between "$\pi$-stacked" porphyrin molecules
and carbon nanotubes was demonstrated very recently
\cite{magadur2008,casey2008}. The main advantage of
"$\pi$-stacking" functionalization is to hardly perturb the
electronic structure of nanotubes (and therefore their optical
and transport properties). In contrast, covalent
functionalization is known to alter the mobility of carriers and
to quench the nanotube luminescence \cite{herranz06}. In the other
hand, "$\pi$-stacking" functionalization performances are quite
poor regarding the yield of functionalization, the
reproducibility and the stability of the suspensions. These
difficulties are due to the weakness of the interactions involved
in "$\pi$-stacking" functionalization and become problematic for
scaling up a controlled process.

In this paper, we report on the synthesis of controlled and
stable SWNT/porphyrin complexes in micelles showing an efficient
energy transfer. The samples are obtained from aqueous
suspensions of individual nanotubes in sodium cholate micelles.
Transmission electron microscopy combined with electron energy
loss spectroscopy demonstrate the presence of porphyrin molecules
in the area of nanotubes. We investigate the functionalization
process by means of optical absorption spectroscopy (OAS) of
suspensions of various concentrations and solvent/water volume
ratios. We point out the crucial role of the organic solvent that
acts as a vector for the insertion of the porphyrin molecules
into the micelle core. Finally, by means of photoluminescence
excitation (PLE) experiments, we demonstrate an efficient energy
transfer from the excited porphyrin to the nanotube.

\part*{Results}

\begin{figure}[h]
    \centering
        \includegraphics*[scale=1]{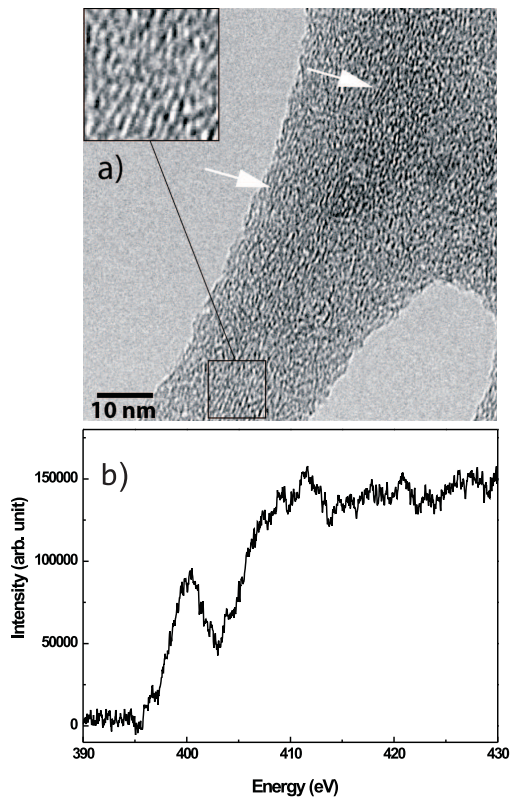}
    \caption{\textbf{TEM and EELS analysis}, a)  TEM images of typical grids prepared from the SWNT/TPP suspensions. The arrows indicate single nanotubes embedded in surfactant. b) EELS spectrum after subtraction of the continuum background, recorded on zone image a).}
    \label{fig:TEM_EELS}
\end{figure}

\section{Structural characterization}
In a first approach, the composition of the SWNT/TPP suspensions
is investigated by TEM.

Figure~\ref{fig:TEM_EELS}~a) shows a typical area where we
identify the presence of individual nanotubes embedded in
surfactant molecules. Nanotubes are difficult to distinguish
since their contrast is shrouded by the amorphous contrast of the
surfactant molecules. However as pointed by the arrows in
figure~\ref{fig:TEM_EELS}~a), one can distinguish the presence of
straight lines with spacing close to 1~nm which is consistent
with the nanotube contrast \cite{springer2006}.

Additional EELS measurements have been recorded on the whole zone
of figure~\ref{fig:TEM_EELS}~a). The corresponding spectrum is
displayed on figure~\ref{fig:TEM_EELS} b). An absorption edge
close to 400 eV typical of the N-K edge is observed indicating the
presence of nitrogen. This element is never observed in pristine
nanotube. Furthermore, the fine structure of the absorption edge
displays two features identified as a $\pi^*$
 peak at 397~eV and a $\sigma^*$
peak at 400~eV. These features are characteristic of nitrogen
bonded to carbon atoms in $sp^{3}$ configuration, which is the
case in the TPP molecule. Since no other component of the
suspension is expected to contain nitrogen, this band is
interpreted as the signature of porphyrin molecules: porphyrin
molecules are present in the very close area of the nanotubes.

\section{Optical properties}

\subsection{Optical absorption}

\begin{figure}[h]
    \centering
        \includegraphics*[scale=.85]{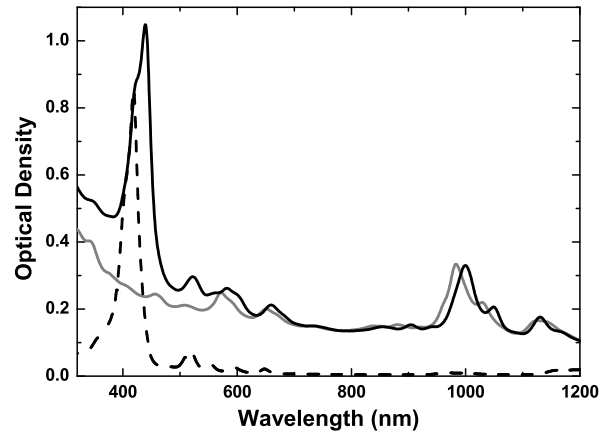}
    \caption{\small \textbf{Optical absorption spectra} of TPP molecules (dashed line, rescaled for clarity), of a reference suspension of SWNT (grey solid
curve) and of SWNT/TPP complexes (black solid curve) all of them
embedded in micelles of sodium cholate in a pH 8 buffer.}
    \label{fig:abs}
\end{figure}

Figure~\ref{fig:abs} displays the OAS of the SWNT/TPP suspension
together with the OAS of a suspension of pure SWNTs and a
suspension of pure TPP, both in SC micelles. The TPP suspension
shows the so-called Soret band at 420~nm and the four weaker Q
bands in the 500-700~nm region.
The OAS of SWNTs consists in a group of lines in the 1000~nm
region corresponding to the so-called $S_{11}$ (lowest)
transitions in semi-conducting nanotubes. Each line in this group
stems from a given $(n,m)$ chiral family. Additional lines near
600~nm and at smaller wavelengths come from $S_{22}$ transitions
superimposed to the onset of $M_{11}$ transitions in metallic
nanotubes. The background in OAS mainly stems from light
scattering and is substracted when analysing the line amplitudes.

The OAS of the SWNT/TPP suspension shows a band at 438~nm with a
shoulder at 420~nm. These lines are attributed to the Soret
absorption band of TPP encased in micelles (at 420~nm, in
agreement with the OAS of pure TPP in micelles) and of TPP stacked
on nanotubes (438~nm). This 18~nm red-shift of the Soret band when
porphyrin molecules are "$\pi$-stacked" on nanotubes is
consistent with previous observations \cite{magadur2008,
cambre2008} and is related to a conformation change of TPP
molecules. The presence of these two bands shows that the
suspension contains both pure TPP molecules encased in micelles
and TPP stacked on nanotubes in micelles. The absorption features
around 1000~nm are consistent with $S_{11}$ absorption lines in
semi-conducting nanotubes. We note however a 7~nm red-shift
compared to the reference SWNT suspension. This bathochromic shift
is due to the presence of TPP close to the nanotubes and to the
resulting interaction.

\subsection{Control of the functionalization}
\begin{figure}[h]
    \centering
        \includegraphics*[scale=.8]{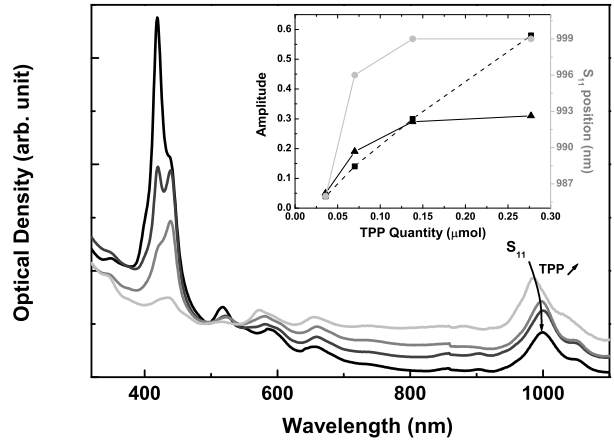}
    \caption{\small \textbf{Functionalization in function of TPP quantity}, Optical absorption spectra of SWNT/TPP complexes for different concentrations of porphyrin with a DCM/water volume ratio of 34\%: 0.04 $\mu$mol (light grey), 0.07 $\mu$mol (grey), 0.14 $\mu$mol (dark grey) and 0.27 $\mu$mol (black). The curves are vertically translated to match at 490~nm (background correction). Inset: Amplitude of the band at 420 nm (dashed line) and at 438 nm (black line) in function of the quantity of porphyrin; Shift of the S$_{11}$ band (grey line) in function of the quantity of porphyrin.}
    \label{fig:fig4_comp_redim}
\end{figure}

In contrast with previous reports on SWNT/TPP stacking, the key of
our work is to provide a reliable method to produce samples of
controlled and reproducible quality. The control of the process is
investigated by studying the functionalization yield as a function
of amount of TPP. Figure~\ref{fig:fig4_comp_redim} shows the OAS
of suspensions made from the same reference SWNT suspension mixed
with a fixed volume of DCM (2.5 mL of SWNT suspension with 0.85 mL
of DCM: volume ratio 34\%, see below) containing a quantity of TPP
ranging from 0.04 $\mu$mol up to 0.28 $\mu$mol. The four curves
are arbitrary shifted (background correction) in order to match
at 490~nm and to facilitate the comparison of the peak amplitudes.
For low TPP concentrations (light grey and grey curves), the band
at 438~nm (stacked TPP) is higher than the one at 420~nm (free
TPP). The two bands have the same intensity for a TPP quantity of
0.14 $\mu$mol (dark grey curve). For a larger amount of TPP, the
band at 420~nm becomes predominant. The amplitude of each band as
a function of TPP concentration are plotted in the inset of
figure~\ref{fig:fig4_comp_redim}. The free TPP band (420~nm) grows
regularly with the initial TPP concentration whereas the
amplitude of the stacked TPP band reaches a plateau above 0.14
$\mu$mol.

These results are interpreted as follows. At low TPP
concentration, TPP molecules are preferentially "$\pi$-stacked"
on SWNTs rather than encased alone in micelles, therefore the
amplitude of the 438~nm band is higher than the one at 420 nm.
For an amount of TPP larger than 0.14~$\mu$mol, no more porphyrin
can stack onto nanotubes and TPP molecules mostly aggregate in
micelles due to the excess of SC: the amplitude of the 438~nm
band saturates while the one at 420~nm grows regularly with the
TPP quantity (inset of figure~\ref{fig:fig4_comp_redim}). The
position of the $S_{11}$ band of SWNTs is another way to track the
stacking of TPP onto SWNTs. For a small amount of TPP, the
$S_{11}$ band (near 1000~nm) is identical to the one in the
reference SWNT suspension: most of the nanotubes are not
functionalized. This band progressively red-shifts with increasing
TPP concentration. Above the saturation threshold ($\sim$ 0.14
$\mu$mol), the position of $S_{11}$ reaches a plateau: most of
the nanotubes are functionalized (grey curve in the inset of
figure~\ref{fig:fig4_comp_redim}).

Importantly, the absorption amplitude of the $S_{11}$ band is
almost identical in both the SWNT/TPP and the reference SWNT
suspensions. This point shows that very few nanotubes are lost in
the process. The increase in TPP quantity only affects the
fonctionalization degree of nanotubes. This is due to the
stability of micelle suspensions of nanotubes. The use of such a
stable starting material is the key for the reproducibility and
control of our method, in contrast to previous attempts where
soluble TPP molecules where used for both solubilization and
functionalization of SWNTs \cite{magadur2008}.

\begin{figure}[h]
    \centering
        \includegraphics*[scale=0.8]{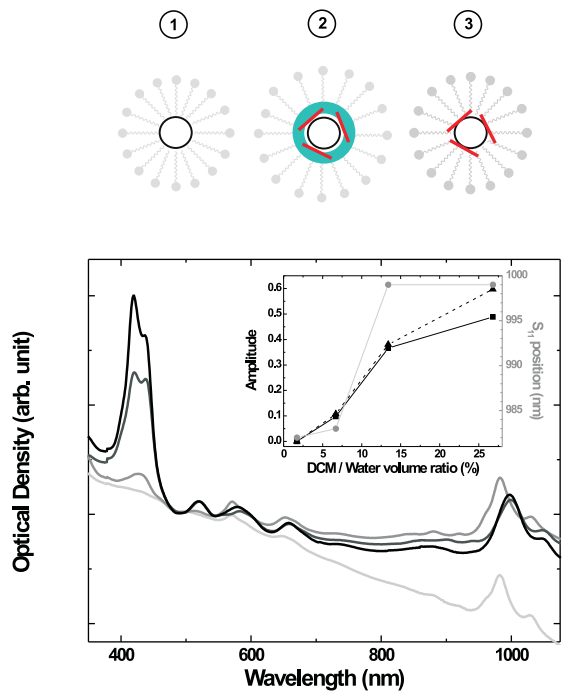}
    \caption{\textbf{Functionalization mechanism}, Top: Scheme of the swelling of the micelle by the dichloromethane (blue) leading to the functionalization of the nanotubes (black line) by the porphyrin molecules (red stick). Bottom: Optical absorption spectra of SWNT/TPP complexes for different DCM/water volume ratios, but with the same TPP mole
    number (0.14 $\mu$mol). Light grey: 2\%, Grey: 7\%, Dark grey: 13\% and Black: 27\%. Inset: Amplitude of the band at 420 nm (dashed line) and at 438 nm (black line) in function of the DCM/water volume ratio; Shift of the S$_{11}$ band (grey line) in function of the DCM/water volume ratio.}
    \label{fig:figdichlo}
\end{figure}

\subsection{Functionalization mechanism}
In view of further developments of this technique, a good
description of the functionalization mechanism is needed. The
SWNT/TPP complex is hydrophobic and is thus expected to stay in
the organic core of the micelles in water suspensions. However,
the formation mechanism of such a complex from pre-existing SWNT
micelles and non water soluble TPP is less straightforward since
TPP molecules have to diffuse through water to reach the core of
the micelles. We show that an organic solvent can act as a vector
and help TPP molecules to penetrate into the micelles. The control
parameter is the DCM/water volume ratio which drives the exchange
area between the two immiscible liquids during the sonication
process. The OAS of SWNT/TPP suspensions for various volumes of
DCM but with constant quantity of TPP (0.14 $\mu$mol) is displayed
in figure~\ref{fig:figdichlo} \cite{note}. For a DCM/water volume
ratio of $\sim$ 2~\% no functionalization is observed (light grey
curve in figure~\ref{fig:figdichlo}). For a volume ratio of
$\sim$~7\%, both the 420 nm and 438 nm Soret bands are observable
but with a weak amplitude (grey curve). For increasing DCM/water
volume ratio up to 27\% (dark grey and black curves), both bands
rise meaning that both the stacking of porphyrin onto nanotubes
and the formation of TPP aggregates in micelles become efficient.
Bearing in mind that the quantity of TPP is kept constant, this
reveals the central role of DCM in the functionalization
mechanism.

Wang \textit{et al.} have shown that water-immiscible organic
solvents swell the micelle hydrophobic core surrounding the SWNT
\cite{wang09}. The authors have mixed a micelle suspension of
nanotubes with immiscible organic solvents. The presence of the
organic solvent inside the micelle is investigated by means of
optical spectroscopy. Typical solvatochromic shifts are observed
in the PL spectra and interpreted as a signature of nanotubes
coated by the solvent. The organic solvent evaporates over time
and after several hours the initial optical spectra are recovered
showing the reversibility of the mechanism. The solvent/water
ratio in Ref.\cite{wang09} is on the order of 50\%, close to the
one used in our studies (34\%) for optimal synthesis conditions.
Therefore we can confidently extend their conclusions to our
work. When mixing the SWNT suspension with the TPP solution, DCM
swells the sodium cholate micelle and therefore acts as a vector
for the interaction between the porphyrin molecules and the
nanotubes. This mechanism is schematically depicted in the upper
panel of figure~\ref{fig:figdichlo} . The starting material is a
nanotube encased in a micelle (step 1). After addition of the DCM
solution of TPP, the DCM swells the hydrophobic core of the
micelle bringing porphyrin molecules in contact with the nanotube
(step 2). As described in reference \cite{wang09}, DCM rapidly
evaporates : TPP molecules stack onto the nanotubes (step 3) and
the new complex remains in the micelle core.

Optical absorption spectra recorded just after and several hours
after the synthesis do not show any noticeable change. We can
therefore conclude that the samples investigated in this study
correspond to step 3 of figure~\ref{fig:figdichlo}.

This micelle core swelling by organic solvents as a way to
enhance the interaction between "$\pi$"-conjugated molecules and
nanotubes is very promising and opens new avenues for nanotube
functionalization.

\section{Stability}
Spectroscopic measurements of SWNT/TPP suspensions were recorded
at different time delays (up to several months) after the
synthesis and no noticeable evolution could be detected. An
important point is that the photoluminescence (PL) signal is
stable for months (see figure S1 in Supplementary Information),
which is the major improvement of our method compared with
previous reports \cite{casey2008, magadur2008}. Furthermore, the
PL intensity of the SWNT/TPP suspension corresponding to the
emission of nanotubes is on the same order of magnitude as the
one of the reference SWNT suspension. It is one order of
magnitude higher than for suspensions obtained with the soluble
TPP stacking method \cite{magadur2008}. In contrast with this
earlier work, this observation confirms that our micelle-based
synthesis ensures that most of the nanotubes remain in the final
product.

\section{Energy transfer}
In this section we probe the interaction between the porphyrin
and the nanotube by means of photoluminescence experiments. The PL
spectra of the SWNT reference suspension and of the SWNT/TPP
suspension are almost identical, except that all bands are
red-shifted by nearly 20~nm as already mentionned for absorption
spectra.

\begin{figure}
\includegraphics*[scale=.6]{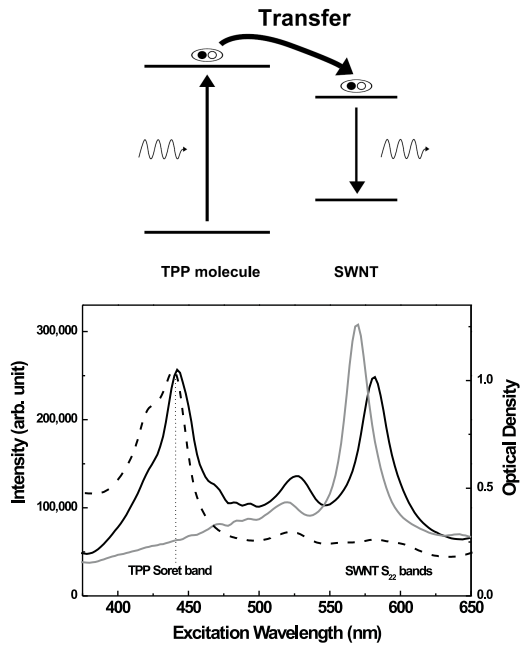}
\caption{\textbf{Energy transfer}, Top: Scheme of the excitonic
levels and of the energy transfer mechanism. Bottom: PLE spectra
of the SWNT suspension (grey) and of the SWNT/TPP suspension
(black) detected at 984 nm and 1007 nm respectively. Dashed line:
optical absorption spectrum of the SWNT/TPP suspension.}
    \label{fig:PLE}
\end{figure}

The PLE spectrum of the reference SWNT suspension detected at
984~nm (figure~\ref{fig:PLE}, grey curve) consists in a broad
line near 560~nm. This couple of PL/PLE wavelengths or
equivalently $S_{11}$ and $S_{22}$ energies allows us to identify
a specific chiral family namely the (6,5) family \cite{bachilo02}.

The PLE spectrum of SWNT/TPP complexes (black curve) strongly
contrasts with the one of the reference SWNT suspension. The PLE
spectrum shows a strong additional line at 441~nm. This line
corresponds to the absorption of stacked TPP (dashed curve). This
PLE peak is the signature of the energy transfer from the
photo-sensitized porphyrin to the nanotube
\cite{casey2008,magadur2008}: when photons are absorbed by a
stacked porphyrin, they give rise to photons emitted by the
nanotube (as depicted in the upper panel of
figure~\ref{fig:PLE})). The line at 580~nm is clearly the
$S_{22}$ line of the nanotubes. Interestingly, this line is
red-shifted of about 10~nm as compared to the reference SWNT
suspension. This result is similar to the shift observed for the
$S_{11}$ lines. Therefore we conclude that this effect arises
from a dielectric screening of excitons and not from a
strain-induced band-shift (as observed for some surfactants
\cite{Kiowski2007,Ohno2007}) that would result in opposite shifts
for $S_{22}$ and $S_{11}$ lines \cite{berger2009}.

Remarkably, the amplitudes of PLE lines corresponding to
porphyrin Soret band and to S$_{22}$ nanotubes transitions are of
the same order of magnitude. This means that the number of photons
emitted by the SWNTs is the same when the SWNTs are excited
directly ($S_{22}$ absorption) or through TPP absorption. We
conclude that the energy transfer from the photo-sensitized
porphyrin molecules to the nanotubes is very efficient.
Nevertheless, further measurements are required in order to
quantify precisely the transfer quantum yield.

In conclusion, we have developed a new method for
"$\pi$-stacking" functionalization of carbon nanotubes by organic
molecules. An organic solvent is used to swell the micelle
surrounding the nanotube and to bring the organic molecules onto
the nanotube in the core of the micelle. This new method allows
the production of controlled, reproducible and stable suspensions
of SWNT/TPP complexes. These complexes show an efficient energy
transfer from the photo-sensitized porphyrin molecules to the
nanotubes confirming the potentialities of this system for
light-harvesting applications. This physico-chemical
functionalization method combines the advantages of
"$\pi$-stacking" functionalization together with a remarkable
stability of the obtained complexes. This is a significant step
towards scaling up a controlled process and designing functional
nano-devices. Finally, this method is neither specific to
nanotubes nor to porphyrins and can therefore be generalized to a
wide range of nano-objects and dye molecules.

\part*{Method}

\textbf{Preparation of SWNT/TPP suspensions:} The nanotubes used
in this study are synthesized by the CoMoCAT process
\cite{resasco1} and produced by SouthWest Nanotechnologies. The
mean diameter of these tubes is about 0.8~nm. The nanotube
suspensions are prepared by adding raw nanotubes at 0.15
mg.mL$^{-1}$ in a pH 8 Normadose buffer (10$^{-2}$ M, Prolabo)
plus 2 wt\% of sodium cholate (Sigma-Aldrich). The mixture is
sonicated for 1h30 with an ultrasonic tip and ultracentrifuged at
120,000 g for 1h00. Then, the supernatant is drawn out. It
consists in a suspension of isolated nanotubes \cite{oconnel02}.

Nanotubes functionalization is achieved by mixing the nanotubes
suspension with a solution of porphyrin in dichloromethane (DCM).
The porphyrin molecules used in this study are
tetraphenylporphyrin (TPP) purified twice by column
chromatography. After adding the TPP solution to the nanotubes
suspension, the mixture is sonicated for 2 hours with an
ultrasonic tip. The sample is placed in a thermostat at
12$^{\mathrm{o}}$C during sonication. The phase corresponding to
DCM in excess is removed and the sample is centrifuged at 3,000 g
for 10 minutes. The supernatant is drawn out and a suspension of
SWNT/pophyrin complexes is obtained.

\textbf{Spectroscopic measurements:} Optical absorption spectra
are recorded with a spectrophotometer (lambda 900 Perkin-Elmer).
A laser diode emitting at 532~nm is used as excitation source for
photoluminescence experiments. The signal is dispersed in a
spectrograph (Spectrapro 2300i, Roper Scientific) and detected by
an IR CCD (OMA V, PI Acton). A UV-VIS Xe lamp and a monochromator
(Spectrapro 2150i, Roper Scientific) are used as tunable light
source for the photoluminescence excitation experiments.
\\
\textbf{TEM measurements:} Transmission electron microscopy (TEM)
images and electron energy loss spectroscopy (EELS) are recorded
using a Libra200 transmission electron microscope at an
accelerating voltage of 200~kV. A  three times diluted suspension
is used for preparing deposits on TEM grids. Then, the grid is
rinsed with water to remove the excess of sodium cholate (SC).


\begin{thebibliography}{10}
\bibitem{note}
\bibinfo{title}{{As shown in the previous section, this amount of TPP is
  sufficient for a full functionalization of all the SWNTs}}.
\end{thebibliography}


\part*{Acknowledgement}

The authors are grateful to D.E. Resasco for providing the CoMoCAT
nanotubes produced by SouthWest Nanotechnologies. LPQM, PPSM and
LPA are "Unit\'es mixtes" de recherche associ\'ees au CNRS
(UMR8537; UMR8531;  UMR8551). This work was supported by the
GDR-E "nanotube" (GDRE2756), grant "C'Nano IdF EPONAD" from
"R\'egion Ile de France" and ANR grant "CEDONA".

\end{document}